\documentclass{IEEEcsmag}

\usepackage[colorlinks,urlcolor=blue,linkcolor=blue,citecolor=blue]{hyperref}
\expandafter\def\expandafter\UrlBreaks\expandafter{\UrlBreaks\do\/\do\*\do\-\do\~\do\'\do\"\do\-}
\usepackage{upmath,color}

\begin{document}

\sptitle{Feature Article: Impact of Google Cookieless solution}

\title{Digital Advertising in a Post-Cookie World: Charting the Impact of Google's Topics API}

\author{Jesús Romero}
\affil{Carlos III University of Madrid, Leganes, (Madrid), 28911, Spain}

\author{Ángel Cuevas}
\affil{Carlos III University of Madrid, Leganes, (Madrid), 28911, Spain}
 
\author{Rubén Cuevas}
\affil{Carlos III University of Madrid, Leganes, (Madrid), 28911, Spain}

\begin{abstract}
Integrating Google's \emph{Topics API} into the digital advertising ecosystem represents a significant shift toward privacy-conscious advertising practices. This article analyses the implications of implementing \emph{Topics API} on ad networks, focusing on competition dynamics and ad space accessibility. Through simulations based on extensive datasets capturing user behavior and market share data for ad networks, we evaluate metrics such as Ad Placement Eligibility, Low Competition Rate, and solo competitor. The findings reveal a noticeable impact on ad networks, with larger players strengthening their dominance and smaller networks facing challenges securing ad spaces and competing effectively. Moreover, the study explores the potential environmental implications of Google's actions, highlighting the need to carefully consider policy and regulatory measures to ensure fair competition and privacy protection. Overall, this research contributes valuable insights into the evolving dynamics of digital advertising and highlights the importance of balancing privacy with competition and innovation in the online advertising landscape.
\end{abstract}

\maketitle


\chapteri{D}igital marketing is a cornerstone of today's Marketing infrastructure, offering unparalleled opportunities for businesses to engage with a vast audience, enhance brand recognition, and drive sales. In the ever-expanding digital landscape, where consumers increasingly allocate their time online, the significance of participating in digital marketing endeavors cannot be overstated.

Yet, traditional digital marketing methodologies, characterized by the use of cookies, have come under intense scrutiny due to their invasive nature, particularly concerning user privacy. This heightened concern has prompted legislative actions such as the General Data Protection Regulation (GDPR) \cite{GDPR} in the European Union and the California Consumer Privacy Act (CCPA) \cite{CCPA}, compelling a reevaluation of advertising practices.

In response to these privacy concerns, digital marketing is witnessing a paradigm shift away from reliance on third-party cookies towards other alternatives. Google's \emph{Topics API} emerges as a notable contender in this landscape, a more privacy-centric solution, offering a mechanism to categorize users based on their inferred interests derived from their browsing activities \cite{Topic_DeveloperChrome, Topic_GitHub}.

While touted as a step forward in safeguarding user privacy, the widespread adoption of \emph{Topics API} raises pertinent questions regarding its ramifications, particularly for the ecosystem of ad networks. These networks, responsible for managing ad spaces across many publishers, are pivotal in facilitating targeted advertising.

This study conducts a comprehensive analysis of the ad network landscape to shed light on the potential implications of the imminent adoption of \emph{Topics API}. Leveraging simulation techniques based on an extensive dataset capturing user interactions across thousands of websites, we evaluate key metrics such as Ad Placement Eligibility, Low Competition Rate, and solo competitor ratio.

Our findings indicate a discernible dichotomy in the impact of \emph{Topics API} adoption on ad networks. Dominant players, exemplified by Google, are poised to reap substantial benefits from this transition, owing to their extensive web presence and infrastructure \cite{MarketShare_Browser, MarketShare_MobileOS, AdNetworksInfo}. Conversely, smaller ad networks with more modest footprints may encounter significant challenges in effectively delivering targeted advertisements in the post-cookie era.

These insights underscore the importance of developing privacy-preserving advertising solutions that uphold user privacy and foster a level playing field within the digital advertising ecosystem, accommodating major stakeholders and smaller players.

\section{BACKGROUND}
\label{sec:background}

The digital marketing ecosystem is a complex landscape focused on efficiently delivering targeted advertisements to users across the web. Over time, the industry has transitioned from traditional direct agreements to programmatic advertising, where ad networks are pivotal in facilitating website buying and selling ad spaces. These ad spaces, defined within web pages by publishers through HTML code snippets, are managed by ad networks, which fill them with relevant ads using various methods such as pre-configured campaigns and real-time auctions. To further optimize the process and promote competition among ad networks while maximizing revenue potential, techniques like header bidding and ad mediation have been developed, enhancing overall efficiency within the ecosystem.

In this ecosystem, Google's \emph{Topics API} presents an innovative approach to privacy-conscious advertising. This proposal aims to replace traditional behavioral advertising, which relies heavily on tracking cookies, with a more privacy-friendly solution. Instead of individualized user attributes, the \emph{Topics API} utilizes coarse-grained interests inferred from users' browsing behavior.

The operation of the \emph{Topics API} involves assigning topics to websites based on their content and tracking users' browsing behavior to infer their interests. Each user is assigned up to five topics weekly, reflecting their browsing habits over the preceding weeks. When a user visits a website with ad spaces, the \emph{Topics API} returns up to three topics, one from each of the previous three weeks, to the ad networks managing those spaces. Ad networks then use this information to deliver ads tailored to users' inferred interests.

While the \emph{Topics API} offers a promising solution for privacy-conscious advertising, its implementation may pose challenges for smaller players in the digital advertising landscape. Accessing relevant topics and effectively competing in the dynamic environment of behavioral advertising requires strategic alignment with broader interest categories.

As we can see, the digital marketing ecosystem continues to evolve, with innovations like Google's \emph{Topics API} reshaping the landscape towards more privacy-conscious advertising practices. Ad networks play a central role in this evolution, optimizing the delivery of targeted ads while navigating the complexities of user privacy and data protection.

\section{METHODS AND DATA}
\label{sec:methods}	

In this study, we aim to assess the ad network access disparity between the \emph{Topics API} and cookie-based methods for behavioral advertising. Using a simulator with real user behavior data, we compare how ad networks access spaces. We aim to evaluate \emph{Topics API} impact on ad networks, validating the hypothesis that larger players will strengthen dominance while smaller ones may exit due to limited topic access. Details on our dataset and simulator follow.

\subsection{Dataset}
\label{subsec:dataset}

The study draws upon Gonzalez et al.'s dataset \cite{data}, which tracks the browsing habits of 1329 users over two months. The dataset comprises 75M connections to 470K websites. Due to privacy constraints, access to the full dataset was restricted. However, the authors provided the necessary information for the research. Specifically, for the simulation, we need the weekly count of unique websites visited by each user and the percentage of revisited websites. These metrics enable the simulation to reflect real browsing behavior. We compute that 
\begin{itemize}
    \item \textbf{Percentile 10} of users visited 33 websites revisiting 14\%.
    \item \textbf{Percentile 25} of users visited 114 websites revisiting 28\%.
    \item \textbf{Percentile 50} of users visited 335 websites revisiting 43\%.
    \item \textbf{Percentile 75} of users visited 668 websites revisiting 58\%.
    \item \textbf{Percentile 90} of users visited 1083 websites revisiting 74\%.
\end{itemize}

\subsection{Simulator}
\label{subsec:simulator}

We have implemented a simulator to emulate the Google \emph{Topics API} ecosystem, including users, websites, and ad networks. The following parameters serve as the elements in configuring simulation executions:

\begin{itemize}
    \item \textbf{Number of Users}: Total participants.
    \item \textbf{Number of Websites}: Total websites.
    \item \textbf{Number of Ad Networks}: Total ad networks.
    \item \textbf{Number of Weeks}: Duration of the simulation in weeks.
    \item \textbf{Pages Per Epoch}: Minimum pages visited weekly.
    \item \textbf{User Loyalty}: Proportion of pages revisited per week.
    \item \textbf{Ads on Site}: Quantity of ad placements.
    \item \textbf{Max Topics}: Maximum topics covered per site.
    \item \textbf{Ad Network Presence}: Proportion of pages with ad network ads.
    \item \textbf{Prop. of Interest Topics}: Percentage of topics ad networks focus on.
\end{itemize}

The simulation involves several steps. First, the number of websites a user will visit is computed. Then, a list of visited pages is generated based on a loyalty parameter and previous calculations. For each visited page, the user registers the website's topics and notes that the ad network on the page has seen them with those topics. The simulation then iterates over the ad spaces on the page, considering different ad networks. It checks if the user has previously shared a topic with the network on the same site and shares topics from the previous epochs. Interested networks are identified based on the shared topics, and one network is randomly chosen as the winner for the ad space. Finally, the top five topics for the current epoch are computed. The outcome of our simulator is the fraction of available ad spaces that each ad network could use to implement behavioral advertising under the \emph{Topics API}.

Our simulations assess how adopting Google's \emph{Topics API} could affect ad networks' access to behavioral ad spaces. By comparing these changes to each network's current market share, we predict whether major players like Google would further solidify their dominance in a post-cookie advertising environment.

It is worth noticing that this simulation is realistic regarding the number of pages with more than one ad network (comparable with the ones using mediation or header bidding nowadays). If anything, there might be more overlap, which will help the small player to have more reach.

\section{THEORETICAL ANALYSIS}

Before simulating real data from advertising networks, we explore the theoretical limits of the Google \emph{Topics API}. We focus on three key variables: the number of ad networks, their presence, and their interest in topics. We conduct two experiments with fixed parameters:

\begin{itemize}
\item \textbf{Number of Users}: 100
\item \textbf{Number of Websites}: 50,000
\item \textbf{Number of Ad Networks}: 50
\item \textbf{Number of Weeks}: 50
\item \textbf{Pages Per Epoch}: 334
\item \textbf{User Loyalty}: 0.43
\item \textbf{Ads on Site}: 10
\item \textbf{Max Topics}: 3
\item \textbf{Ad Network Presence}: 0.8
\item \textbf{Proportion of Interest Topics}: 1
\end{itemize}

In our first experiment, we examined how varying the number of ad networks (from 10 to 200) and their presence (1\% to 100\%) affects ad space utilization. We found that with a significant presence of ad networks (40\% or more), just ten networks could fill all ad spaces effectively using behavioral advertising. However, major platforms like Google Ads, LinkedIn, and Bing had presence levels below 20\%. Conversely, scenarios with numerous small ad networks (below 2\% presence) left a notable portion of ad spaces unfilled, with 100 networks filling 82\% and 200 networks filling 96\%.

Our study highlights Google's \emph{Topics API} limitations with many small ad networks and emphasizes the importance of ad network presence across websites for competitive ad space acquisition. Smaller to medium-sized ad networks with limited website presence (5\% or less) may struggle to secure ad spaces for behavioral advertising.

In our second experiment, with 50 ad networks, we varied their interest in topics (10\% to 100\%) while considering different presences (1\% to 100\%).

Similar to the first experiment, ad network presence emerged as crucial. Scenarios with less than 10\% presence resulted in unclaimed ad spaces unless ad networks engaged with at least 80\% of the topics. With ad networks engaging only 10\% of topics, significant challenges arose under the Google Topics API, especially with a limited website presence. Even with 10\% presence, 37\% of ad spaces remained vacant, and with 40\% presence, 4\% of spaces were still vacant.

These findings underscore the necessity for ad networks to establish a broad presence across websites and topics to compete effectively within the Google \emph{Topics API} framework. This may disadvantage niche ad networks catering to specific domains, as the API seems to favor major players with extensive website presence and topic coverage.

\section{RESULTS}
\label{sec:results}	

In this section, we analyze the effect of the \emph{Topics API} on ad networks' performance, focusing on competition reduction, sole competitor prevalence, and behavioral ad targeting constraints. Understanding these dynamics is crucial amid the API's widespread adoption.

We aim to uncover relationships between ad network presence, competition, and behavioral targeting efficacy. Throughout this section, when we mention an ad network's inability to serve ads, we refer specifically to its incapacity for behavioral ads.

We run 11,000 simulations using the following simulation parameters:

\begin{itemize}
    \item \textbf{Number of Users}: 10,000
    \item \textbf{Number of Websites}: 50,000
    \item \textbf{Number of Ad Networks}: 174
    \item \textbf{Number of Weeks}: 55
    \item \textbf{Pages Per Epoch}: 334
    \item \textbf{User Loyalty}: 0.43
    \item \textbf{Ads on Site}: 10
    \item \textbf{Max Topics}: 3
    \item \textbf{Ad Network Presence}: Varied based on market share
    \item \textbf{Proportion of Interest Topics}: 1
\end{itemize}

For each ad network, we measure:

\begin{itemize}
	\item \textbf{Low Competition Ratio}: Percentage of ad spaces where an ad network faces fewer competitors for behavioral advertising than the total number of ad networks on the website.
	\item \textbf{Sole Competitor Ratio}: Percentage of ad spaces where an ad network faces no competition for behavioral advertising.
	\item \textbf{Ad Placement Eligibility}: Percentage of ad spaces where an ad network qualifies to display behavioral ads.
\end{itemize}

To conclude this section, we will verify whether the obtained results are consistent across users' behaviors. To do this, we also ran 5,000 simulations with different configurations for each of the percentile groups of users described in the section dataset.

\subsection{Low Competition Ratio}

\begin{figure}[ht]
	\centering
	\includegraphics[width=.5\textwidth]{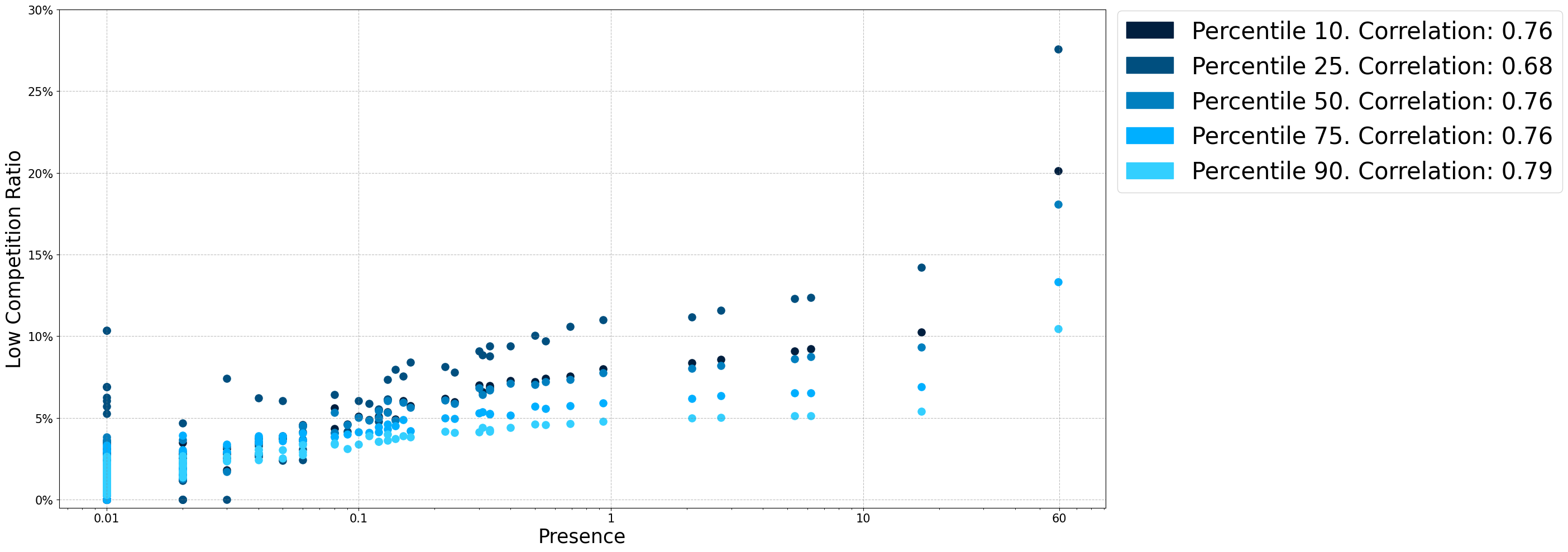}
	\caption{On the figure, can be seen five plots that refer to user groups' percentiles 10, 25, 50, 75, and 90 of visited pages. Each dot represents an ad network with its presence and the ratio of ad spaces with fewer competitors than possible. The respective Spearman's correlations are 0.76, 0.68, 0.76, 0.76, 0.79}
	\label{fig:UserComparison_PresenceVSLessCompetitantsRatio}
\end{figure}

In this section, we will focus on the median user group presented in Figure  \ref{fig:UserComparison_PresenceVSLessCompetitantsRatio} to explore how the presence of ad networks correlates with the level of competition in ad spaces.

The data reveals a clear trend: as an ad network becomes larger, the more ad spaces it holds a competitive edge. This trend is particularly relevant for larger ad networks, which consistently have a high percentage of ad spaces facing less competition.

Major ad networks, characterized by their extensive market share and widespread presence, enjoy a significant advantage in this regard. They tend to occupy a large portion of ad spaces with limited competition, giving them better chances of winning bids and displaying their ads effectively. With more ad spaces experiencing reduced competition, these networks can strategically secure prime ad placements, thereby increasing their visibility and the likelihood of ad displays.

Among the prominent ad networks, Google AdSense (with a presence of 59.24\%) stands out. 18.02\% of the ad spaces associated with Google AdSense benefit from a significant competitive advantage, facing limited competition from other ad networks. This underscores Google AdSense's strong position in the advertising landscape, potentially leading to higher ad revenues and attracting more advertisers seeking optimal ad placements.

These findings emphasize the crucial role of ad network size and presence in shaping competition dynamics within the advertising domain. The dominance of major ad networks in securing ad spaces with reduced competition not only reflects their extensive inventory but also suggests potentially lower costs for occupying these spaces, as fewer competing ad networks will be bidding.

\subsection{Sole competitor}
\label{section:Having_no_competition}

\begin{figure}[ht]
	\centering
	\includegraphics[width=.5\textwidth]{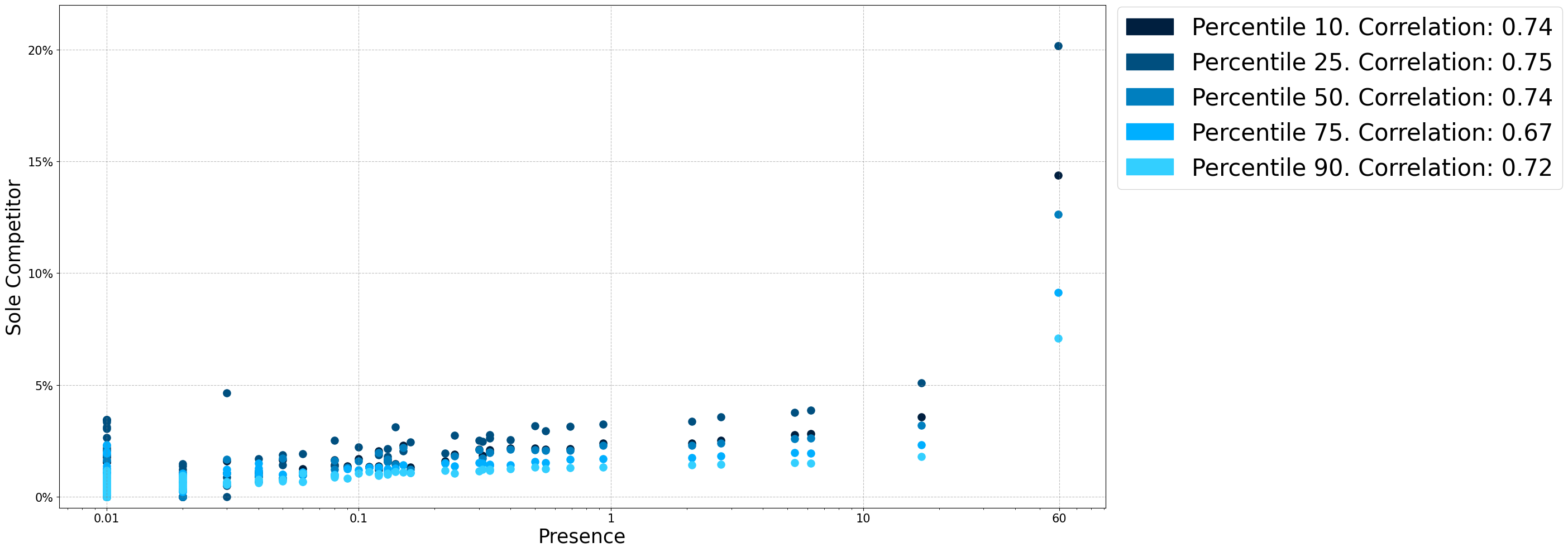}
	\caption{On the figure, can be seen five plots that refer to user groups' percentiles 10, 25, 50, 75, and 90 of visited pages. Each dot represents an ad network with its presence and the ratio of ad spaces where they were the only competitors. The respective Spearman's correlations are 0.74, 0.75, 0.74, 0.67, 0.72}	
	\label{fig:UserComparison_PresenceVSOnlyCompetitantsRatio}
\end{figure}	

Now that we have an understanding of the Low Competition Ratio, we can gain a further understanding of the competition dynamics by examining the sole competitor analysis of the median users in figure \ref{fig:UserComparison_PresenceVSOnlyCompetitantsRatio}. Larger ad networks, with more presence, tend to dominate this category by occupying more ad spaces where they face no competition from other ad networks. This underscores their ability to secure advertising opportunities exclusively.

Remarkably, the data indicates that only the larger (1\% presence or more) ad networks manage to secure the sole competitor in more than 2\% of ad spaces. Conversely, a majority (60\%) of ad networks struggle to secure even 0.05\% of ad space without competition, highlighting the considerable challenge smaller ad networks face.

Among these larger players, Google AdSense stands out. It manages to be the sole competitor in an impressive 12.59\% of ad spaces, solidifying its dominance in the advertising market. This exclusive access to behavioral advertising in numerous ad spaces gives Google AdSense a significant competitive advantage, attracting advertisers seeking maximum exposure and engagement.

These findings underscore the difficulties smaller ad networks encounter in accessing exclusive ad placements and emphasize the need for innovation and alternative strategies to remain competitive. Moreover, the dominance of major ad networks in securing these spaces warrants ongoing scrutiny and potential regulatory intervention to ensure fair competition and a level playing field.

\subsection{Ad Placement Eligibility}

\begin{figure}[ht]
	\centering
	\includegraphics[width=.5\textwidth]{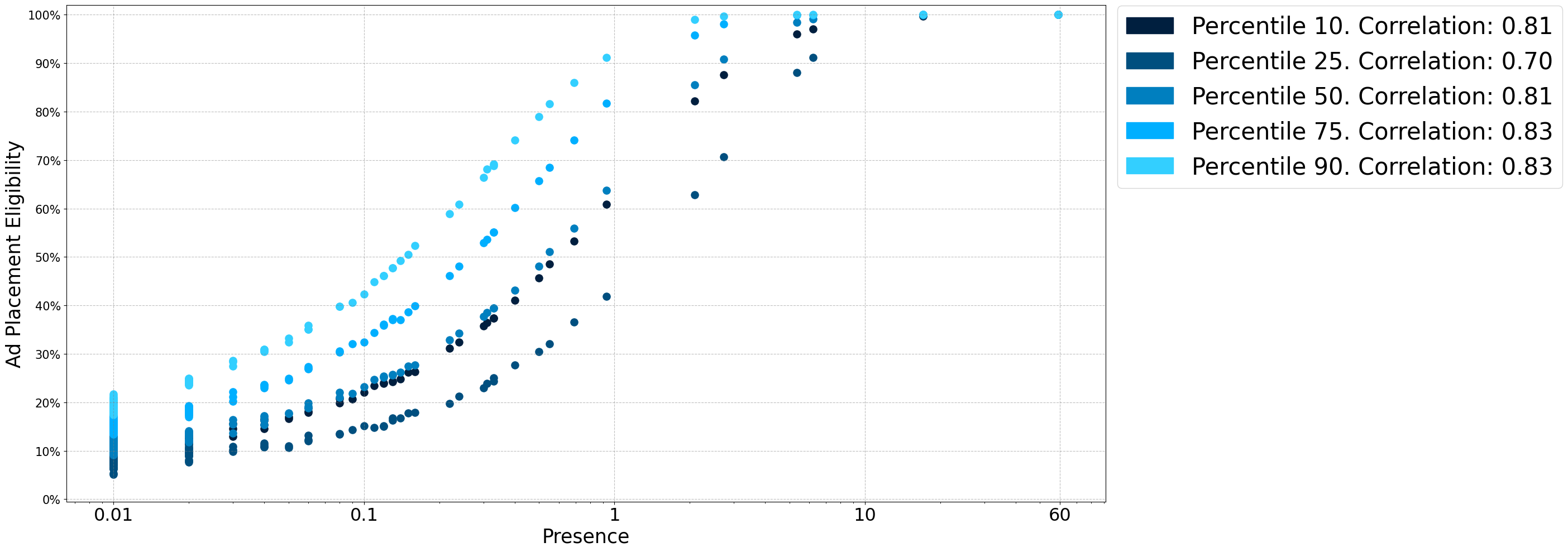}
	\caption{On the figure, can be seen five plots that refer to user groups' percentiles 10, 25, 50, 75, and 90 of visited pages. Each dot represents an ad network with its presence and the ratio of ad spaces where they could compete. The respective Spearman's correlations are 0.81, 0.70, 0.81, 0.83, 0.83}
	\label{fig:UserComparison_PresenceVSCompetitantsRatio}
\end{figure}

The central focus of this section lies on the median users of figure \ref{fig:UserComparison_PresenceVSCompetitantsRatio} to analyze Ad Placement Eligibility. Upon examination, it becomes evident that the ad networks with the biggest presence exhibit an outstanding track record, achieving close to 100\% coverage or even attaining a full 100\% coverage. This implies they can serve ads in almost all spaces without significant limitations. In other words, there are very few instances, if any, where they will lose business opportunities for being unable to display relevant behavioral advertisements.

On the contrary, the smaller ad networks, representing most of the sample, face a significant challenge. Our analysis reveals that smaller players, with a presence below 0.03\%, only receive a topic for advertisement placement in less than 15\% of the ad spaces. They represent 77\% of the ad networks, that a staggering 85\% of the times, cannot do behavioral targeting on the available ad spaces. As a result, they miss out on potential revenue generation opportunities and face a substantial barrier in capitalizing on their available advertising inventory.

The glaring contrast between large and small ad networks in serving behavioral ads directly impacts revenue generation, causing smaller networks to suffer significant losses. Their inability to capitalize on most ad spaces puts them at a competitive disadvantage, jeopardizing their financial viability in the evolving landscape of \emph{Topics API} and ad serving. Consequently, many smaller ad networks may face considerable challenges in surviving and thriving, emphasizing the urgent need for innovative strategies to enhance their relevance, topic coverage, and overall competitiveness in this new era of ad serving.

\subsection{Generalisation to different user behaviors}

In this section, we aim to demonstrate that the conclusions drawn in previous sections remain consistent across various user behaviors. To do this, we classify users into five percentiles (10, 25, 50, 75, and 90) based on their behavior, as explained in the dataset section.

We present our results in three key figures: Figure \ref{fig:UserComparison_PresenceVSLessCompetitantsRatio}, Figure \ref{fig:UserComparison_PresenceVSOnlyCompetitantsRatio}, and Figure \ref{fig:UserComparison_PresenceVSCompetitantsRatio}. Each figure comprises five plots corresponding to the five user behavior groups.

As anticipated, our findings confirm that the conclusions derived from earlier sections hold true across different user behaviors. Regardless of the number of pages visited or user loyalty, dominant ad networks with substantial market shares tend to maintain their competitive edge. They consistently exhibit a higher ratio of ad spaces where they can engage in bidding and secure favorable ad placements with less competition. Additionally, they are more likely to be the sole competitors in various ad spaces, further consolidating their market position. Spearman's correlations support these findings, with the lowest correlation coefficient at 0.67, while being 0.74 or higher for 11 out of 15 plots.

These findings raise concerns about potential monopolistic tendencies in the market and emphasize the importance of thoughtful policy and regulatory measures to ensure fair competition and a level playing field for all ad networks, irrespective of their size or market share.

\section{CONCLUSION}

\subsection{Impact on ad networks}

Integrating the \emph{Topics API} into the advertising ecosystem significantly impacts ad networks. Larger networks gain competitive advantages, monopolizing ad spaces and potentially charging higher prices. This limits revenue opportunities for smaller networks, threatening their viability and diversity in ad content. Smaller networks struggle to secure ad spaces and face increased competition. To compete, they must innovate. Overall, the \emph{Topics API} presents challenges for ad networks, particularly smaller players, requiring policy interventions to foster fairness and diversity. Further research is needed to understand evolving dynamics and inform strategic decisions.

\subsection{Google}

The elimination of third-party cookies might boost user privacy, but Google's decision isn't purely altruistic. By controlling the ecosystem, Google can gain a competitive advantage over other ad tech companies and increase its market share.

The small percentage of ad spaces in which Google will have no competition may not seem significant, but it is important to remember that we are talking about trillions of ad spaces worldwide. Even a tiny percentage of ad spaces could represent a substantial number of ad spaces, and therefore revenue, for Google. The global advertising market was worth \$825.86 billion in 2022 \cite{Globalad87:online}, and internet users worldwide are around 5.18 billion as of April 2023 \cite{Internet37:online}. Out of all of this, Google has a presence of 60\%. In the worst case (90 percentile), they are securing 7.1\% of their ad spaces. This means that they are securing 4.2\% of all the ad spaces where only Google can do behavioral advertising.

It's worth noting that Google has faced criticism and fines in recent years for violating privacy and competition regulations. For example, in 2019, Google was fined \$170 million by the Federal Trade Commission (FTC) for violating children's online privacy rules \cite{Googleguilty45:online}. In February 2023, the U.S. Justice Department sued Google for Monopolizing Digital Advertising Technologies \cite{Officeof80:online}. In January 2022, regarding Cookieless, the UK's Competition and Markets Authority launched an investigation of suspected anti-competitive conduct by Google. \cite{Investig85:online}. The last update, on January 2024, is that the CMA investigation has multiple competition-related concerns in Google's efforts to eliminate third-party cookies \cite{CMAQ420247:online}. With increasing regulatory scrutiny, Google must demonstrate that it prioritizes users.

Our final question would be whether this has been done deliberately? Google's proposed solution will help the users' privacy, but while doing that, it seems Google will reinforce its dominant position in the market. 

\section{ACKNOWLEDGMENTS}
The research leading to these results received funding from the European Union's Horizon 2020 innovation action program under the grant agreement No 871370 (PIMCITY project); the Ministerio de Econom\' {i}a, Industria y Competitividad, Spain, and the European Social Fund(EU), under the Ram\' {o}n y Cajal program (Grant RyC-2015-17732);  the Ministerio de Ciencia e Innovaci\' {o}n under the project ACHILLES (Grant PID2019-104207RB-I00); the Community of Madrid synergic project EMPATIA-CM (Grant Y2018/TCS-5046); and the Fundaci\' {o}n BBVA under the project AERIS.

\def\refname{REFERENCES}
\bibliographystyle{IEEEtran}
\bibliography{art_cookieless}


\end{document}